\documentclass[runningheads]{llncs}
\usepackage[T1]{fontenc}
\usepackage{graphicx}
\usepackage{hyperref}
\begin{document}
%
\title{Particle Builder - Learn about the Standard Model while playing against an AI\thanks{This demo has been accepted for presentation at the AIED 2025 Interactive Events Track.}}

%
%
\author{Mohammad Attar \inst{1} \and Andrew Carse \inst{1} \and Yeming Chen  \inst{1} \and Thomas Green \inst{1} \and Jeong-Yeon Ha \inst{1} \and Yanbai Jin \inst{1} \and  Amy McWilliams \inst{1} \and Theirry Panggabean \inst{1} \and Zhengyu Peng \inst{1} \and Lujin Sun \inst{1} \and Jing Ru \inst{1} \and Jiacheng She \inst{1} \and Jialin Wang \inst{1} \and Zilun Wei \inst{1} \and Jiayuan Zhu \inst{1} \and Lachlan McGinness \inst{1,2} \orcidID{0000-0002-3231-4827}}
\institute{Australian National University \and
Commonwealth Scientific and Industrial Research Organisation
\email{lachlan.mcginness@anu.edu.au}}

\authorrunning{M. Attar, et al.}
%

%
\maketitle              
\begin{abstract}
Particle Builder Online is a web-based education game designed for high school physics students. Students can play against an AI opponent or peers to familiarise themselves with the Standard Model of Particle Physics. The game is aimed at a high school level and tailored to the International Baccalaureate and the Australian Curriculum. Students from four schools in Canberra took pre/post-tests and a survey while completing a lesson where they played Particle Builder. Students' understanding of particle physics concepts improved significantly. Students found the game more enjoyable and effective than regular classroom lessons.

\keywords{Physics Education Research, Standard Model of Particle Physics, Educational AI, Games}

\end{abstract}

\vspace{-1.0cm}
\section*{Demo Links}
\vspace{-0.35cm}
The game can be accessed online and there are also videos which introduce and explain aspects of the game. A four minute video which gives an introduction to the game, with little explanation of the underlying physics can be found here: \url{https://youtu.be/5bBMdEUQj_w} . An extended version (six minutes) of the demonstration with some of the basic physics principles introduced is here: \url{https://youtu.be/zQ-N7aOE3P4}. This link navigates to the online game \url{https://particle-builder.anu.edu.au/} and for reference we also provide  the physical board game, hosted by CERN, can be downloaded here: \url{https://zenodo.org/records/14210858}

\vspace{-0.20cm}
\section*{System Presentation} 
\vspace{-0.25cm}
\subsection*{Originality and Strengths}
\vspace{-0.15cm}
High school curricula often include the Standard Model of Particle Physics but lack corresponding interactive learning resources \cite{KranjcHorvat2022What}. Particle Builder extends analogy based resources to cover many different concepts in a single activity \cite{McGinness2019ThreeD}, \cite{McGinness2019Printable}, \cite{Gettrust2010Quark} \cite{Strunk2018Model}.

Particle Builder Online's originality lies in its application of interactive game mechanics (card comparison, strategic placement, annihilation) directly mapping to complex Standard Model concepts within an accessible web-based platform. The inclusion of an AI opponent allows for repeatable, engaging practice. The system addresses a documented lack of interactive resources for this topic.

Initial classroom studies indicate the approach is effective, showing significant learning with an average gain of 0.18 and a normalised gain of 0.23. The survey results also show high student engagement compared to traditional lessons with students rating on average $6.3$ out of 7 for enjoyment and $5.4$ for learning where $4$ means comparable to a normal lesson. 

\subsection*{Technical Specifications and Gameflow}
Particle Builder is hosted on a DVM12 Apache Server using Nginx. It is a Unity-based game made with C\#, ShaderLab and HLSL. The game was heavily adapted from a commercial template. The system includes a tutorial mode guiding users through mechanics and physics principles, a primary mode allowing play against an AI opponent, and a peer-to-peer mode.

The tutorial mode is designed to teach students how to play the game. The students are guided through a deterministic path of the appropriate level and the physics of the game is explained while they work through the different phases of the game. 

In the AI mode students start choose a level, a particle system and then they begin a game. Students take turns with the AI, to draw a card which contains information about a particle from the Standard Model which they compare to a hidden opponent's card. Students are advantaged by knowledge of the particles contained in the cards and learn while playing. The game ends when either the AI or the student completes their particle system.

Although at the time of writing the human vs. human mode is only available on the Australian National University campus, the team intends to extend this for players anywhere in the world soon. The game flow then is very similar to playing against the AI. 

\subsection*{AI Implementation}

The AI system in Particle Builder employs a deterministic rule-based approach with strategic randomisation for educational efficacy. During gameplay, the AI can perform a range of actions including playing cards onto board slots or target slots, using an ability (annihilation, discarding or attracting opponent particles), to moving cards already on the board, initiating card comparisons and choosing to keep or swap cards with its opponent. 

To inform its decisions, the AI utilizes a utility called `GetAllSlots' to analyse the current board state. This is used by the AI game logic to assess the possibility of potential moves. 

The AI is intentionally designed to support the game's educational goals rather than optimize for winning. Since students typically play only a few times to learn particle physics concepts, a highly optimized AI could be counterproductive to the learning experience. Therefore, the AI does not employ complex lookahead strategies. The AI follows a simple hierarchy of rules: it prioritizes annihilating opponent target cards, then completing its own target, then attacking opponent cards, and finally placing or discarding the current card.

In specific situations requiring choices less critical to the core placement strategy, the AI incorporates randomness. For instance, selecting which card statistic to use during the comparison phase, or deciding whether to swap cards after winning a comparison, is randomised. This prevents the AI from having an inherent advantage based on game knowledge that a new player lacks and introduces unpredictability.

A potential area for future development is a “hard mode,” where players could challenge a more sophisticated AI. This version could implement a minimax search algorithm to evaluate the best move by exploring potential game states a fixed number of turns into the future. While such a system may increase engagement for some learners, it is unlikely to enhance the educational outcomes of the game and may even discourage new learners.

\begin{credits}
\subsubsection{\ackname} We would like to thank Harri Leinonen who worked with Lachlan McGinness to create the first version of Particle Builder. We also thank and acknowledge Rowan McGinness who developed all the art used in the game.

The ethical aspects of this research have been approved by the ANU Human Research Ethics Committee (Protocol 2024/0566).

We would like to thank CSIRO and the ANU TechLauncher program for their funding and support of this work.
\end{credits}
%
%
%
%

\end{document}